\documentclass[%
 reprint,
superscriptaddress,
 amsmath,amssymb,
 aps,
]{revtex4-2}

\usepackage{graphicx}% Include figure files
\usepackage{dcolumn}% Align table columns on decimal point
\usepackage{bm}% bold math
\usepackage{hyperref}% add hypertext capabilities
\usepackage{xcolor}
\begin{document}

\preprint{APS/123-QED}

\title{SINAPSE: A lightweight deep learning framework for accurate and explainable neutron-$\gamma$ discrimination}% Force line breaks with \\

\author{Thomas~Carreau}
 \email{carreau@lpccaen.in2p3.fr}
 \affiliation{University of Caen Normandy, ENSICAEN, CNRS/IN2P3, LPC Caen UMR6534, F-14000 Caen, France}%

\author{Adrien~Matta}%
 \email{matta@lpccaen.in2p3.fr}
\affiliation{University of Caen Normandy, ENSICAEN, CNRS/IN2P3, LPC Caen UMR6534, F-14000 Caen, France}%

\author{Owen~Syrett}
\affiliation{CEA, DAM, DIF, Arpajon, F-91297, France}%
\affiliation{Université Paris-Saclay, CEA, Laboratoire Matière en Conditions Extrêmes, Bruyères-le-Chatêl, F-91680, France} 

\author{Beno\^it~Mauss}
\affiliation{CEA, DAM, DIF, Arpajon, F-91297, France}%
\affiliation{Université Paris-Saclay, CEA, Laboratoire Matière en Conditions Extrêmes, Bruyères-le-Chatêl, F-91680, France} 

\author{David~Etasse}%
\affiliation{University of Caen Normandy, ENSICAEN, CNRS/IN2P3, LPC Caen UMR6534, F-14000 Caen, France}%

\author{Cyril~Lenain}
\affiliation{CEA, DAM, DIF, Arpajon, F-91297, France}%
\affiliation{Université Paris-Saclay, CEA, Laboratoire Matière en Conditions Extrêmes, Bruyères-le-Chatêl, F-91680, France} 

\author{Pierre~Morfouace}
\affiliation{CEA, DAM, DIF, Arpajon, F-91297, France}%
\affiliation{Université Paris-Saclay, CEA, Laboratoire Matière en Conditions Extrêmes, Bruyères-le-Chatêl, F-91680, France} 

\author{Julien~Taieb}
\affiliation{CEA, DAM, DIF, Arpajon, F-91297, France}%
\affiliation{Université Paris-Saclay, CEA, Laboratoire Matière en Conditions Extrêmes, Bruyères-le-Chatêl, F-91680, France} 

\author{David~Regnier}
\affiliation{CEA, DAM, DIF, Arpajon, F-91297, France}%
\affiliation{Université Paris-Saclay, CEA, Laboratoire Matière en Conditions Extrêmes, Bruyères-le-Chatêl, F-91680, France}

\author{Patrick~Copp}
\affiliation{Los Alamos National Laboratory, Los Alamos, New Mexico 87545, USA}
\author{Matthew~Devlin}
\affiliation{Los Alamos National Laboratory, Los Alamos, New Mexico 87545, USA}
\author{Charl\`ene~Surault}
\affiliation{CEA, DAM, DIF, Arpajon, F-91297, France}%
\affiliation{Université Paris-Saclay, CEA, Laboratoire Matière en Conditions Extrêmes, Bruyères-le-Chatêl, F-91680, France} 
\author{Jason~Surbrook}
\affiliation{Los Alamos National Laboratory, Los Alamos, New Mexico 87545, USA}
%}%

\date{\today}% It is always \today, today,

\begin{abstract}
Traditionally, neutron-$\gamma$ discrimination in organic scintillators relies 
on techniques such as time-of-flight (ToF) selection and pulse-shape 
discrimination (PSD). However, particle identification through graphical cuts 
remains challenging in the low-charge regime due to poor signal-to-noise ratios 
(SNR).
In this work, we propose SINAPSE, a lightweight deep learning framework for
accurate and explainable neutron-$\gamma$ discrimination in the low-charge 
regime. The framework employs a dual-branch architecture that combines a 
1-dimensional convolutional autoencoder for waveform denoising with a 
classifier for particle identification.
Random augmentations are applied to high-SNR waveforms to simulate low-charge 
conditions, enabling robust extrapolation into regimes where conventional PSD 
labels are unreliable.
We show that SINAPSE achieves superior denoising performance compared to 
conventional digital signal processing techniques, and outputs well-calibrated 
probabilities, consistent with traditional graphical cuts.
Finally, we apply SHAP (SHapley Additive exPlanations) values to show that 
model decisions are driven by physically meaningful pulse-shape features, 
confirming consistency with established PSD principles.
\end{abstract}

\maketitle

%\tableofcontents

\section{\label{sec:intro}Introduction}

Several liquid organic scintillator arrays have been developed for the measurement of prompt fission neutrons (PFN)~\citep{haight2012two} and beta-delayed neutrons~\citep{martinez2014monster}. Liquid organic scintillators have long demonstrated their effectiveness for fast-neutron detection owing to their fast timing characteristics and their ability to discriminate neutrons from $\gamma$ rays using pulse-shape discrimination (PSD) at sufficiently large pulse heights.
Neutron-$\gamma$ discrimination techniques are essential across a wide range of
scientific and industrial fields such as radioprotection, nuclear measurements,
and physics experiments.

Traditionally, neutron-$\gamma$ discrimination in organic scintillators relies
on techniques such as time-of-flight (ToF) selection and pulse-shape
discrimination.
In a typical fission reaction both neutrons and $\gamma$ rays are emitted and 
travel at substantially different speeds. In an experiment, one can then 
identify almost only $\gamma$ rays by selecting events around the prompt
$\gamma$-ray peak in the ToF distribution.
In the case where ToF information is unavailable,
or when $\gamma$-ray can be detected within the same time-window as neutrons, 
PSD techniques are necessary to discriminate between the detection of either 
particle. For example, a fraction of $\gamma$ rays are emitted
with delays ranging from a few nanoseconds to more than hundreds of nanoseconds following fission.
On average, a few percent of $\gamma$ rays are emitted belatedly because of long-lived isomers populated during the fission fragment de-excitation~\citep{talou2023nuclear}, while some $\gamma$ rays are emitted following neutron induced reactions on the structural materials of experimental setups.
The impossibility to discriminate the particle types by ToF motivates the 
improvement of PSD performances with the development of detection capabilities, 
electronics, and data analysis.
 
PSD exploits differences in the scintillation decay components induced by neutrons and $\gamma$ rays in organic scintillators. 
The scintillation process comprises short and delayed decay-time
components~\citep{brooks1979development}. The relative proportion of each component depends on whether a $\gamma$ ray or a neutron was
detected. Typical PSD techniques use different integration gates to
highlight these ratios. Each pulse is reduced into a PSD factor that depends on whether a neutron or a $\gamma$ interacted with the liquid scintillator.
The PSD factor is defined in our case as the ratio of the charge integrated
over the prompt peak, $Q_\mathrm{short}$, to the charge integrated over a wider 
time window that captures the full scintillation decay, $Q_\mathrm{long}$. 
When this PSD factor is plotted as a function of the total integral of the
signal, see Fig.~\ref{fig:psd}, two distinct clusters emerge. 
The neutron events form the lower band, reflecting their larger delayed 
scintillation component.
At low light outputs, the spread of the PSD factor increases as the 
signal-to-noise ratio decreases, and $\gamma$-ray and neutron response signals 
cannot be fully distinguished anymore.

To address this limitation, recent studies have investigated the use of machine learning and deep learning approaches for neutron-$\gamma$ discrimination~\citep{fabian2021artificial,garnett2024neutralnet,panda2025discrimination}. By learning discriminative features directly from waveform data, these models have demonstrated improved performance over traditional PSD methods, particularly in the low-energy regime where handcrafted features become unreliable.

In this work, we propose a lightweight deep learning framework, SINAPSE (\emph{SImulateur Numérique pour l’Apprentissage Profond sur Système Embarqué}), for
accurate and explainable neutron-$\gamma$ discrimination in the low-charge
regime. The description of the experimental setup is provided in section~\ref{sec:exp}. Our dataset preparation strategy is described in section~\ref{sec:dataset}. The architecture and training details are given in section~\ref{sec:method}. Finally, our results are presented in section~\ref{sec:results} and model explainability is explored in section~\ref{sec:interpretability}.

\section{\label{sec:exp}Experimental setup}

\begin{figure}[t!]
    \centering
    \includegraphics[width=1\linewidth]{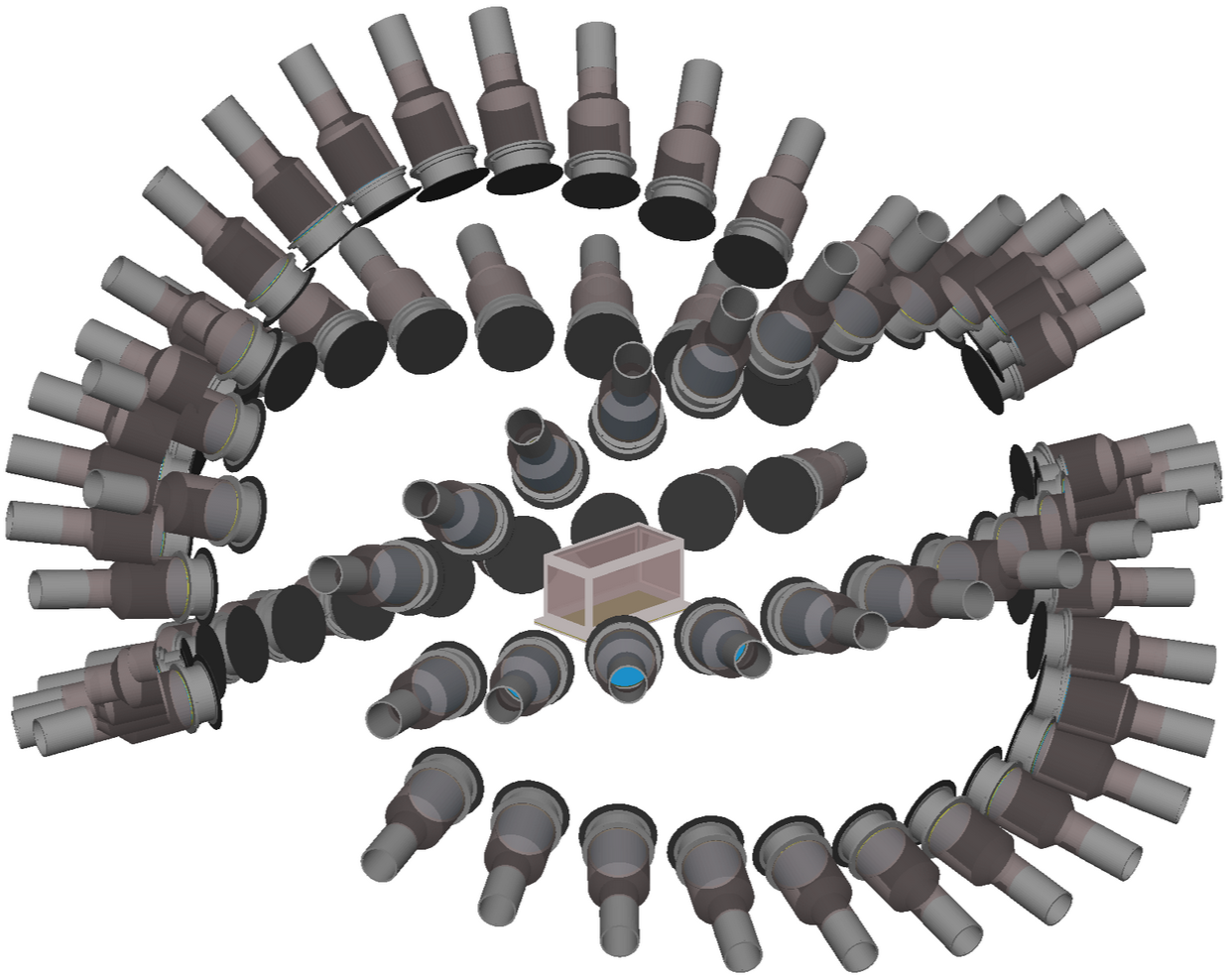}
    \caption{Visualization of the prompt fission neutron (PFN) spectra measurement setup, as simulated in Geant4~\citep{agostinelli2003geant4} with \emph{nptool}~\citep{matta2016nptool}.}
    \label{fig:setup}
\end{figure}

The experimental data for the training and evaluation of our framework were
recorded with the VENDETA array, in its prompt fission neutron spectra configuration, described in~\cite{syrett2025vendeta} and represented in Fig.~\ref{fig:setup}

\subsection{Detectors}

The VENDETA array is composed of 72 EJ-309 type liquid scintillator cells.
Each cylindrical cell is 50 mm in height and 127 mm in diameter and is connected directly to a R4144 PhotoMultiplier Tube (PMT) from Hamamatsu. The PMT’s
output was connected to a custom made dual gain amplifier. First, the signal
was shaped and the rise time was increased from about 2 ns to 6 ns, to ensure a sufficient number of sampling points for timing measurement by the FASTER DAQ
system (see section~\ref{subsec:faster}) with its 500 Megasamples per second sampling frequency. Then,
the signal was split into two outputs which amplify the PMT signal’s amplitude,
by about a factor 10 for the high gain (HG) output, and by about 1 for the
low gain (LG) output. Thus, the HG stage was sensitive to low
amplitude signals while the LG stage could still collect higher amplitude signals
avoiding the saturation of the digitizer.

The array surrounded a fission chamber~\citep{laurent2021new} that contained
a $^{240}\text{Pu}$ spontaneous fission source. The $^{240}\text{Pu}$ was
electro-plated by the JRC-Geel onto 12~\textmu m thick aluminium backings acting as cathodes in a stack of ionization chambers. The fission chamber
was used as a source of neutrons and $\gamma$ rays, and provides the detection 
of $\alpha$ particles and fission fragments. The charge deposit measurement 
allows for $\alpha$-fission discrimination, while the time measurement gives the 
start corresponding to the occurrence of a fission event. The fission event 
detection time was compared to detection times in VENDETA for ToF measurements.

In the current configuration the array was placed at the Weapon Neutron
Research (WNR) facility of the Los Alamos Neutron Science Center (LANSCE)
on the 15° left flight path (4FP15L).

\subsection{FASTER electronics}\label{subsec:faster}

The experiment utilized the FASTER (Fast Acquisition SysTem for nuclEar
Research)~\citep{faster}, a modular digital acquisition system designed for 
nuclear physics experiments.
Specifically, the CARAS daughter cards were employed, which feature a 500
Megasamples per second, 
12-bit low-noise analog-to-digital converter (ADC) for high-fidelity signal 
digitization. Each CARAS card supports two channels with a dynamic range of
$\pm 1.15 \ \mathrm{V}$, and an adjustable input offset between -1.1 V and 
1.1 V. The analog bandwidth is optimized to 100 MHz via a passive 
low-pass filter, ensuring an excellent signal-to-noise ratio for signals with 
rise times greater than 3~ns.

The CARAS cards were mounted on SYROCO AMC C5 motherboards, which provide 
synchronization across all boards within a microTCA crate. This setup enables 
precise time alignment and data aggregation, crucial for achieving high 
temporal resolution.
The sampler mode of the FASTER system was used to record full waveforms, 
allowing for offline processing. Post-acquisition, a constant fraction 
discriminator (CFD) algorithm~\citep{knoll2010radiation} was applied to the 
sampled signals to optimize timing resolution, while pulse integrals were 
calculated using integration gates adjusted so as to maximize the figure of 
merit (FoM) for neutron-$\gamma$ discrimination.

\section{\label{sec:dataset}Datasets}

The performance of deep learning models for neutron-$\gamma$ discrimination critically depends on the quality and representativeness of the training data. In this section, we describe the $^{240}\text{Pu}$ spontaneous fission dataset used in this study, including its statistical properties, event selection criteria, and preprocessing pipeline.

\subsection{Dataset statistics}

Our dataset contains 2{,}484{,}531 signals associated to spontaneous fission 
events from $^{240}\text{Pu}$, of which 45\% have been labelled either as
neutron or $\gamma$ ray using time-of-flight and pulse shape discrimination 
techniques. Each signal is recorded as a 200 ns waveform sampled at 100 points. 

\begin{figure}[t!]
    \centering
    \includegraphics[width=1\linewidth]{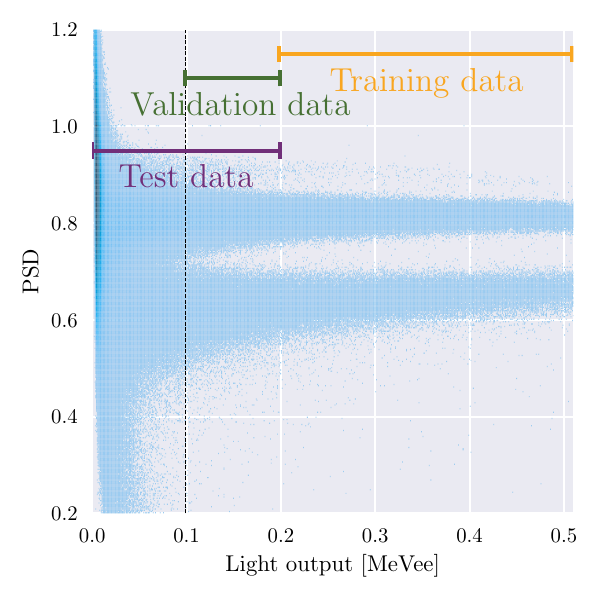}
    \caption{Density matrix of the pulse shape discrimination factor
      $(\text{PSD}=\frac{Q_\mathrm{short}}{Q_\mathrm{long}})$ as a
    function of light output. Data above 0.2 MeVee are used for
    training. Validation data are selected between 0.1 and 0.2 MeVee. 
  Extrapolation is carried out below 0.1 MeVee.}
    \label{fig:psd}
\end{figure}

In particular, a subset of $\gamma$-ray signals can be unambiguously selected by gating the $\beta$ 
distribution around the prompt $\gamma$-ray peak located at $\beta = 1$, where 
$\beta$ is defined as
the ratio of the particle velocity to the speed of light in vacuum. PSD is 
further required to differentiate prompt fission neutrons
from $\gamma$ rays arriving within the same time frame. Among the
1{,}117{,}702 annotated signals, 23.6\% have been identified as neutrons. 
As illustrated in Fig.~\ref{fig:psd}, the PSD factor for
neutrons depositing little energy in the VENDETA detectors partly mixes with
$\gamma$ rays, preventing reliable identification with graphical cuts.  

\subsection{Event selection}

\begin{figure}[b!]
    \centering
    \includegraphics[width=1\linewidth]{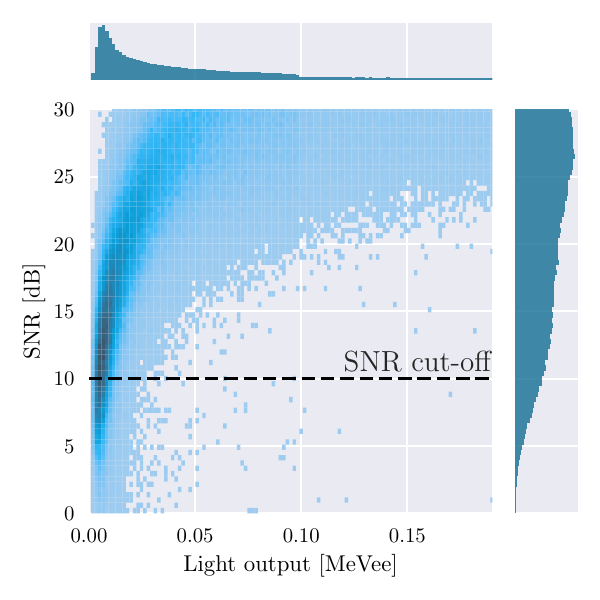}
    \caption{Signal-to-noise ratio as a function of light output for the test set waveforms. Only waveforms with a signal-to-noise ratio above the horizontal dashed line (SNR cut-off) are kept within the test set. The one-dimensional marginal distributions for the light output and the signal-to-noise ratio are also represented in the top and right panels.}
    \label{fig:snr_cutoff}
\end{figure}

Our event selection strategy is presented in
Table~\ref{table:event_selection}.
All signals for which the signal-to-noise ratio (SNR) is below 10 dB, at which
signal information is washed out by noise, are excluded from the dataset. The
SNR is defined as
\begin{equation}
  \text{SNR}_\mathrm{dB} =
  20\log_{10}\left(\frac{A_\text{signal}}{\sigma_\text{noise}}\right),
\end{equation}
where $A_\text{signal}$ is the RMS of the signal and $\sigma_\text{noise}$ is
that of the noise, which is equivalent to the standard deviation since the noise
has zero mean. The SNR as a metric for waveform quality has several advantages: \emph{(i)} it is less dependent on the specific experimental apparatus, \emph{(ii)} by using an SNR cut-off rather than a light output threshold, we are able to preserve signals down to the hardware electronic threshold of 6 keVee as shown in Fig.~\ref{fig:snr_cutoff}.
In this work, the PMT detected charge is converted in electron-equivalent light output units (keVee or MeVee), accounting for the quenching effect. The quenching effect is a reduction in the scintillation light yield due to the high ionization density. A light output of 1 keVee (or MeVee) is equivalent to the light output of an electron depositing 1 keV (or 1 MeV) of energy in the scintillator.
Events with a total light output below 0.1 MeVee, for which reliable 
identification via PSD is not achievable, are excluded from the training and 
validation data.
Event selection is therefore based on both the integrated charge and the 
$\beta$ distributions. Charge-based selection is defined as
follows: events with light output values between 200 keVee and 510 keVee
are assigned to the training set, those with light outputs between 100 keVee
and 200 keVee are split between the validation and test sets using stratified 
sampling by label, and events with light output below 100 keVee 
are added to the test set.
For the $\beta$ distribution, a range of $0 < \beta < 0.6$ is selected to 
exclude prompt $\gamma$ events ($0.8 < \beta < 1.2$), which are retained as an 
additional test subset. 
Finally, we perform undersampling of the $\gamma$ events, which form the majority class, on 
the training and validation sets so as to work with balanced splits. 
After applying these criteria, the dataset comprises 78{,}596 prompt 
$\gamma$ events, 100{,}000 waveforms in the training set (50\% $\gamma$, 50\%
neutrons), 60{,}000 in the validation set (50\% $\gamma$, 50\% neutrons), and 
293{,}821 in the main test set, the majority of which remain unlabelled 
(12\% $\gamma$, 43\% neutrons, 45\% non-annotated). 

\begin{table*}[t!]
\centering
\begin{tabular}{l l l l}
\hline
\textbf{Subset} & \textbf{Light output range [keVee]} & \textbf{$\beta$ range} & \textbf{Size} \\
\hline
Train 
& 200--510
& 0--0.6
& 100{,}000 (50\% $\gamma$, 50\% neutrons) \\

Val 
& 100--200
& 0--0.6
& 60{,}000 (50\% $\gamma$, 50\% neutrons) \\

Test 
& 6--200
& 0--0.6
& 293{,}821 (12\% $\gamma$, 43\% neutrons) \\

Prompt $\gamma$ 
& 6--100 
& 0.8--1.2
& 78{,}596 $\gamma$ \\
\hline
\end{tabular}
\caption{Event selection criteria.}
\label{table:event_selection}
\end{table*}

\subsection{Signal normalization and random augmentations}

\begin{figure*}[t!]
    \centering
    \includegraphics[width=1.0\textwidth]{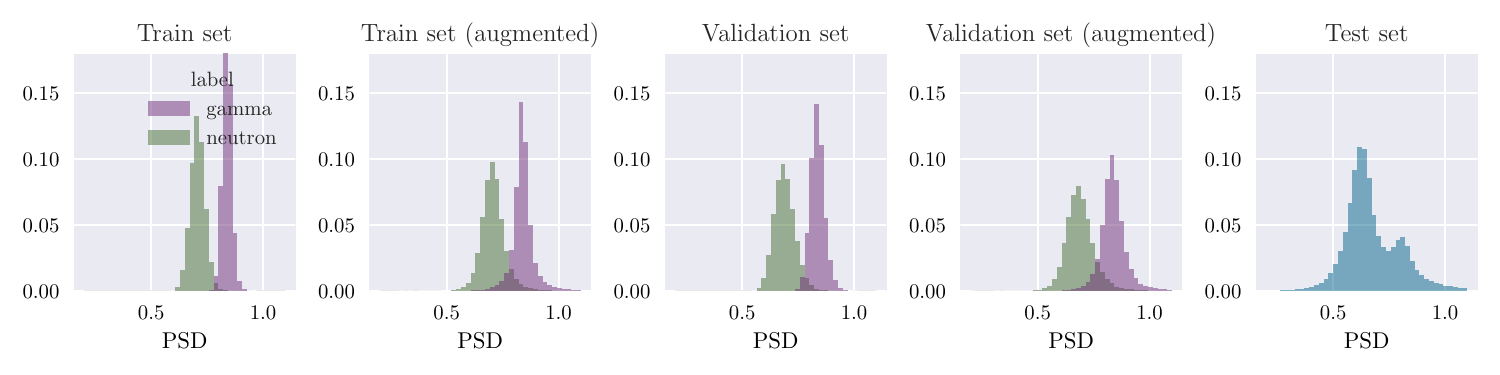}
    \caption{Pulse shape discrimination factor distributions for the train set
      and validation set before and after augmentation, and the test set. Only 
    events characterized by light output values between 6 and 100 keVee are
  considered for the test set.}
    \label{fig:psd_splits}
\end{figure*}

The signals from the training set, perfectly discriminated between neutron and
$\gamma$ rays, are normalized and then transformed through random
augmentations in order to mimic walk effects and poor SNR observed in the low charge regime. Thus, model robustness is improved and reliable extrapolation to the low-charge regime is enabled.

Random augmentations are performed using the following preprocessing pipeline.
First, the input waveform undergoes baseline correction and $z$-score normalization, which consists in converting signal samples $x_i$ into standard scores $z_i$ by
\begin{equation}
z_i = \frac{x_i - \mu}{\sigma},
\end{equation}
where $\mu$ is the mean waveform amplitude and $\sigma$ is the standard deviation.
A random temporal shift is then applied to the signal, with a shift value uniformly 
sampled between -4 and 4 ns, to account for the walk effect. 
This effect arises because pulses from neutron or gamma events of 
different energies have different amplitudes. Indeed, with a fixed-threshold
discriminator, higher pulses cross the threshold earlier, resulting in a 
systematic timing offset. Note that the walk effect is partly mitigated when 
using a constant fraction discriminator, which is the case in this experiment.
Subsequently, to train the model to denoise low SNR signals, we create our targets by 
adding Gaussian white noise to the shifted copy. The noise variance is drawn
uniformly at random, such that the resulting SNR falls between 10 dB and the
original SNR of the clean signal.
Fig.~\ref{fig:psd_splits} shows the PSD factor distributions for the different
data splits. We can see that our augmentation strategy allows us to recreate
the mixing between neutron and $\gamma$ clusters encountered at low light 
output values from signal characterized by high SNR values. This is validated 
by fitting two Gaussian distributions, corresponding to the neutron and $\gamma$
clusters, then computing the figure of merit defined as
\begin{equation}
  \text{FoM} = \frac{|\mu_n - \mu_\gamma|}{\sigma_n + \sigma_\gamma},
\end{equation}
where $\mu_n$ ($\mu_\gamma$) and $\sigma_n$ ($\sigma_\gamma$) are the mean and
standard deviation of the neutron ($\gamma$) distribution. 
This quantity reflects the amount of mixing of the clusters. Thus, the 
augmented data used for training ($\mathrm{FoM} = 1.8$) is produced with a 
level of mixing similar or superior to the data for which we aim to improve 
discrimination (test set has $\mathrm{FoM} = 1.6$).

\section{\label{sec:method}Method}

This section presents our dual-branch framework, which combines a 1-dimensional convolutional autoencoder for waveform denoising with a parallel classifier for particle identification. We employ multitask learning to jointly optimize reconstruction and classification, while exploring variational Bayesian last layer for improved probability calibration.

\begin{figure*}[t!]
    \centering
    \includegraphics[width=0.9\linewidth]{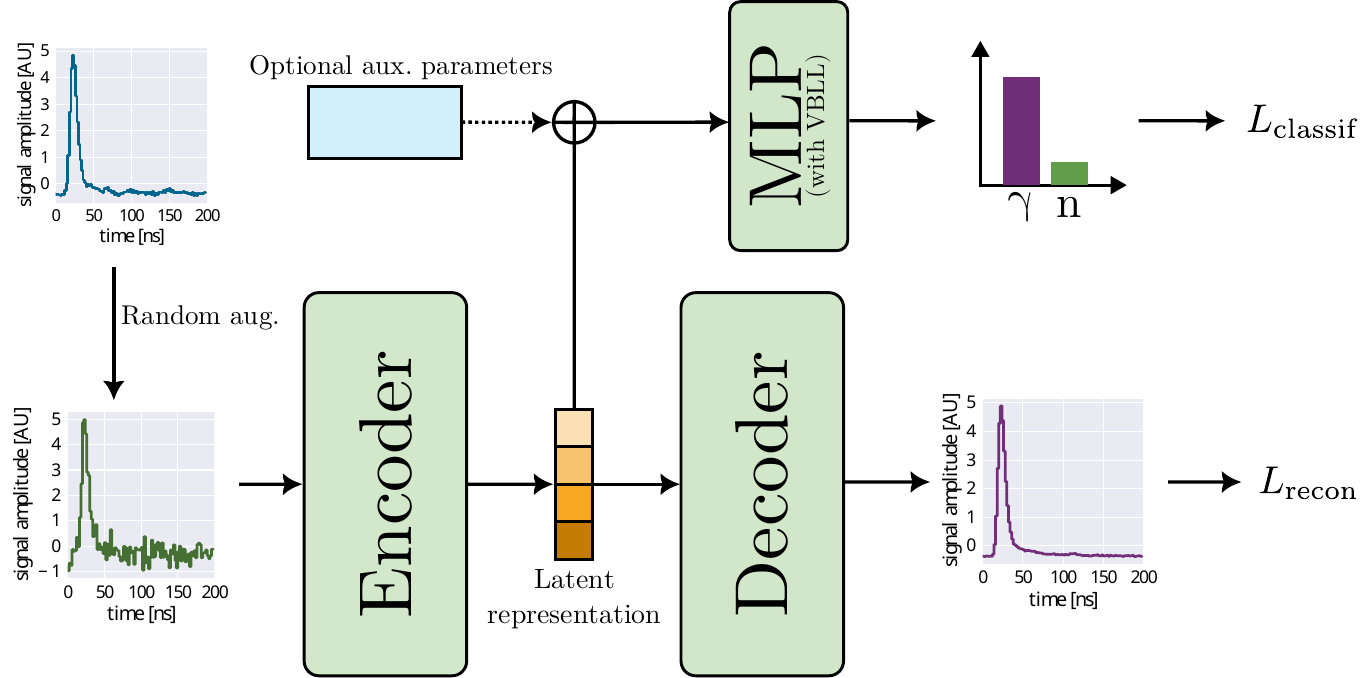}
    \caption{Overview of the SINAPSE framework. An input waveform is first
    augmented using random transformations (temporal shift and Gaussian noise).
  The augmented signal is then processed by an encoder block (1D CNN) to
extract informative features. The resulting embedding is sent in two parallel
branches: a decoder block, mirroring the encoder, for reconstructing the denoised signal, and an
multilayer perceptron (MLP) either with a final fully-connected layer or a 
variational Bayesian last layer for classification. 
Optional auxiliary parameters can be concatenated to the embedding along the 
channel dimension before being passed to the classifier. Finally, 
classification and reconstruction losses are computed by the discrepancy 
between predicted and their corresponding ground truths.}
    \label{fig:sinapse_framework}
\end{figure*}

\subsection{Architecture}

An overview of the framework is shown in Fig.~\ref{fig:sinapse_framework}. The 
encoder receives an augmented 1D waveform as input and extracts a compact set 
of informative features. It consists of a simple lightweight 1D convolutional 
neural network (CNN) composed of two convolutional layers with 16 and 32 output 
feature maps, respectively, each using a kernel size of 3 and a stride of 1, 
followed by ReLU activation functions. A final fully connected (FC) layer maps 
the convolutional output to an 8-dimensional latent representation, 
corresponding to a compression ratio of 12.5. This embedding is then fed into 
two parallel branches: a decoder that reconstructs the denoised waveform, and 
a classifier responsible for particle identification. The decoder mirrors the 
encoder and consists of two 1D transposed convolution layers.
Our classifier consists of two linear layers with a ReLU non-linearity in 
between.
We also experiment with a variational Bayesian last layer
(VBLL)~\citep{harrison2024vbll}.
Replacing the traditional final fully-connected layer with a VBLL is expected
to improve predictive accuracy as well as to get calibrated uncertainty 
estimates without adding computational overhead. 
More specifically, we employ the discriminative classification model, denoted 
in the following as D-VBLL. This model corresponds to a multinomial Bayesian 
logistic regression on learned features, implemented as a Bayesian last-layer 
classifier that integrates a Gaussian prior over the last-layer weights within 
the VBLL framework to capture uncertainty while performing standard 
discriminative classification.
While Bayesian last layer models consider only the uncertainty over the output 
layer of the network, they have been shown to perform comparably to more 
complex Bayesian models~\citep{watson2021latent}.
Note that auxiliary parameters, for example the time-of-flight, can optionally 
be concatenated to the latent vector along the channel dimension before being 
passed to the classifier. 
Based on our experiments, we do not find that adding $\beta$ as an auxiliary
parameter yields significant improvements in neutron-$\gamma$ discrimination.
Still, concatenating energy and ToF information to the embedding is expected to 
improve performance in multiclass pulse-shape-analysis-based 
particle identification on GRIT data~\citep{grit}.

\subsection{Training}

The network is trained in a multitask fashion, the shared encoder being
optimized simultaneously for waveform reconstruction and particle
classification. This joint training strategy encourages the latent 
representation to capture both the global pulse morphology relevant for 
denoising and the subtler shape differences required for neutron–$\gamma$
discrimination.

For the reconstruction branch, we employ a mean-squared-error (MSE) loss
$\mathcal{L}_\mathrm{MSE}$ between the decoded waveform and the clean target 
signal.
For the classification branch, we use the binary cross-entropy loss
($\mathcal{L}_{\mathrm{cls}} = \mathcal{L}_\mathrm{BCE}$) for the model with a final 
fully-connected layer and the evidence lower bound (ELBO) loss for the model 
with a VBLL following~\cite{harrison2024vbll} 
($\mathcal{L}_{\mathrm{cls}} = \mathcal{L}_\mathrm{ELBO}$).
The ELBO encourages the model to balance accurate label prediction with 
calibrated uncertainty by combining a data-fit term with a Kullback–Leibler 
divergence regularizer on the variation last-layer weights. The total training 
objective is defined as a weighted sum of the reconstruction and classification 
losses,

\begin{equation}
  \mathcal{L} = \lambda_\mathrm{rec} \mathcal{L}_\mathrm{MSE} +
  \lambda_\mathrm{cls} \mathcal{L}_\mathrm{cls},
\end{equation}

\noindent with weights tuned so as to balance the magnitudes of the two terms during 
early training.

We train the model for up to 3,000 epochs using a batch size of 2,048.
Optimization is performed with the AdamW optimizer using a learning rate 
of $3\times 10^{-4}$ and weight decay of $10^{-4}$. 
A cosine learning-rate schedule with linear warm-up is applied. The warm-up 
stabilizes the early training dynamics (first 50 epochs), while the cosine 
decay improves convergence by smoothly reducing the learning rate over time. 
Early stopping with a patience of 300 epochs is employed, using the
classification loss on the validation set as the stopping criterion. 
The model and training procedure are implemented in 
\texttt{PyTorch}~\citep{paszke2019pytorch}.
Training requires approximately two hours on a single NVIDIA RTX A1000 GPU, 
including data preprocessing steps.

\section{\label{sec:results}Experiments and results}

To evaluate the performance of our proposed framework, we conduct a systematic assessment of both its signal denoising capabilities and particle identification accuracy. This section presents quantitative metrics and qualitative analyses across multiple experiments, including comparisons with classical denoising methods and evaluations of model robustness under varying noise levels and training set sizes.

\subsection{Signal denoising}

\begin{figure}
    \centering
    \includegraphics[width=1.0\linewidth]{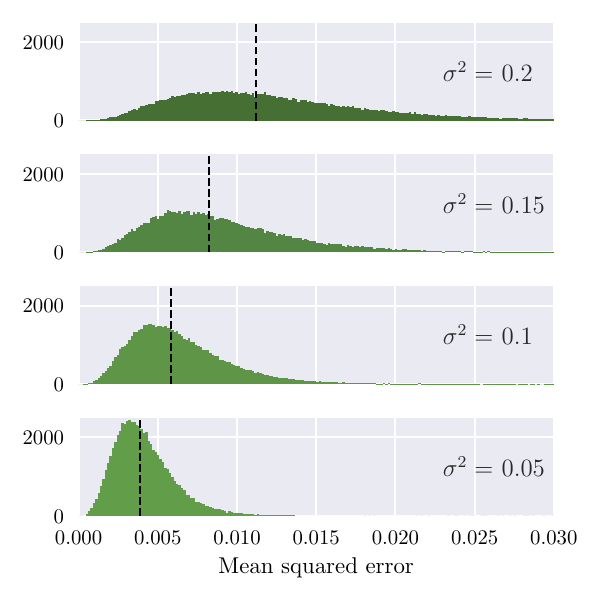}
    \caption{Distribution of per-signal mean squared error between augmented
      then denoised signals and original, normalized signals for four selected variance values
    for the added Gaussian noise: 0.05, 0.1, 0.15, and 0.2. Vertical dashed lines indicate median values.}
    \label{fig:mse_vs_var}
\end{figure}

We quantitatively assess the denoising capabilities of our model by
constructing an independent dataset comprising signals characterized by light
output values greater than 100 keVee and add Gaussian noise with zero mean and
variance ranging from 0.05 to 0.2 to normalized signals.
Fig.~\ref{fig:mse_vs_var} shows the distribution of per-signal mean squared
error (MSE) between augmented then denoised signals and original signals for four noise levels: $\sigma^2=0.05$, $\sigma^2=0.1$, $\sigma^2=0.15$, and $\sigma^2=0.2$. It is seen that the distribution flattens  and that the median MSE value increases as we increase the variance. A very good reconstruction is obtained, with the median MSE ranging from $3.9 \times 10^{-3}$ for $\sigma^2 = 0.05$ to $1.1 \times 10^{-2}$ for $\sigma^2 = 0.2$.

\begin{figure*}[t!]
    \centering
    \includegraphics[width=0.9\linewidth]{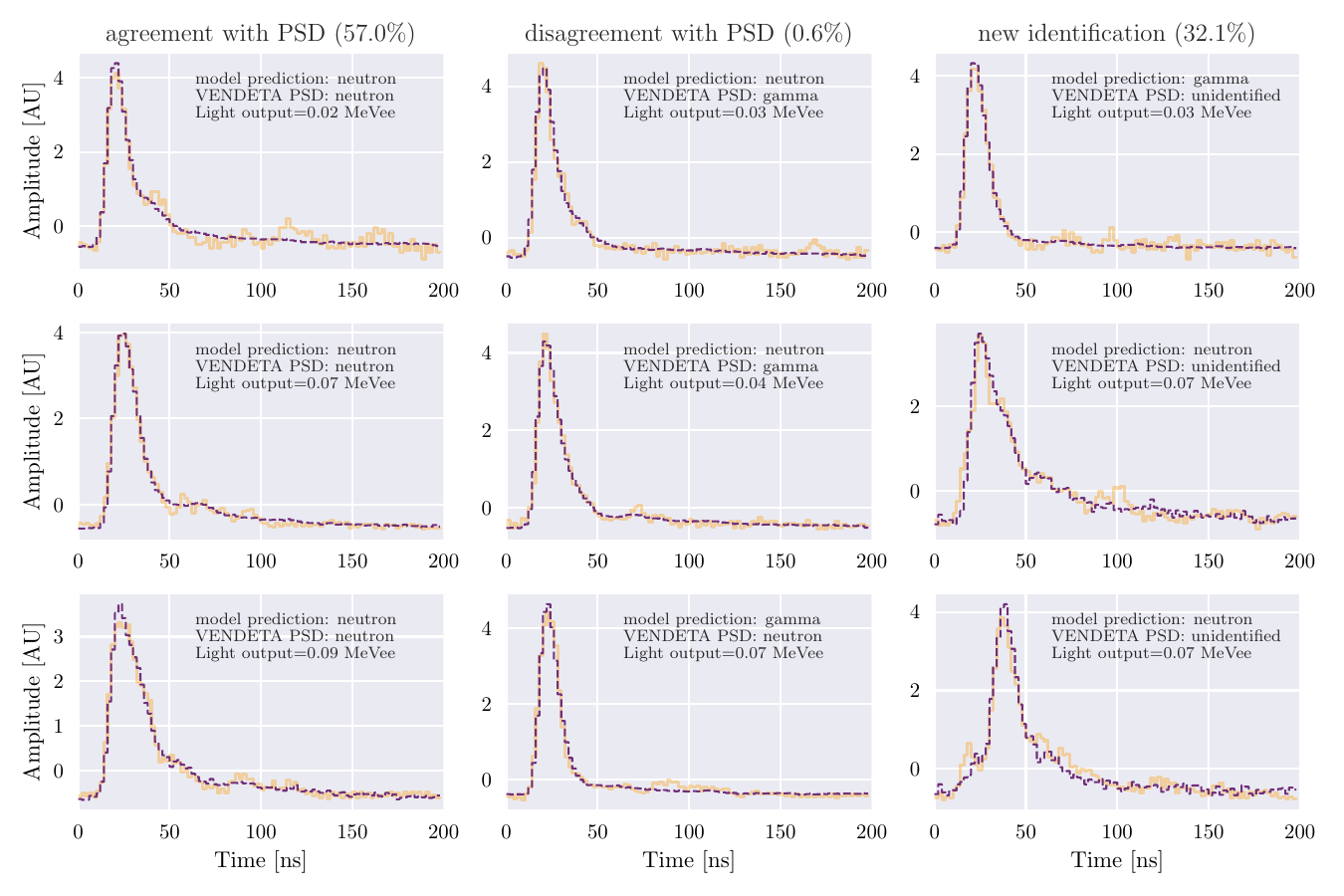}
    \caption{Left: Example signals for which model predictions are in agreement
    with VENDETA PSD results. Middle: Example signals for which model
  prediction are in disagreement with VENDETA PSD results. Right: Example
signals for which identification is permitted by our model but not by the
traditional PSD method. Solid orange lines represent input signals with light
output below 100 keVee and dashed purple lines represent corresponding denoised signals.}
    \label{fig:tp_fp_unidentified}
\end{figure*}

Qualitative examples of the reconstructed signals of the test dataset are presented in 
Fig.~\ref{fig:tp_fp_unidentified}. As illustrated, the model performs robust 
noise suppression while preserving the characteristic pulse shape. Three 
representative cases are shown: events for which the model prediction agrees 
with the PSD label (57.0\% of the test dataset, left column), events where the model 
prediction disagrees with the PSD label (0.6\%, middle column) and events that
cannot be distinguished with the PSD technique, but were confidently given a
class by the model (32.1\%, right column).
Here, we define a neutron as any event with a predicted probability greater than 0.9, 
and a $\gamma$ as any event with a predicted probability less than 0.1. This leaves us with 10.3\% of signals for which we do not provide a label.
These examples highlight the model’s ability not only to denoise the waveform, 
but also to provide a particle identification in cases where graphical cuts are 
inconclusive as detailed in subsection~\ref{subsec:id}.

\begin{figure}
    \centering
    \includegraphics[width=\linewidth]{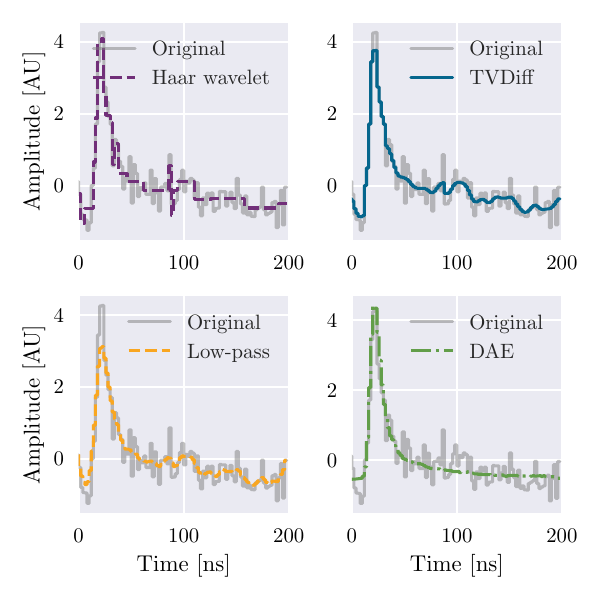}
    \caption{Examples of signal denoising using four selected approaches: Haar
      wavelet decomposition~\citep{haar1911theorie},
      total-variation denoising~\citep{rudin1992nonlinear}, low-pass filtering, and our denoising
  autoencoder (DAE).}
    \label{fig:dsp}
\end{figure}

A comparison with classical denoising algorithms is provided in 
Fig.~\ref{fig:dsp}. More specifically, we qualitatively benchmark the denoising
capabilities of our best performing model (bottom-right panel) against Haar wavelet 
decomposition~\citep{haar1911theorie}, total-variation (TV) 
denoising~\citep{rudin1992nonlinear}, and low-pass filtering.
Among the classical approaches, the Haar transform offers the best 
compromise, effectively removing high-frequency noise while maintaining the 
peak amplitude. In contrast, TV denoising tends to suppress sharp features due 
to its intrinsic objective of penalizing differences between consecutive 
samples; even after testing several regularization strengths, the 
best-performing setting still
noticeably alters the pulse maximum. Overall, these comparisons demonstrate 
that the proposed model achieves superior denoising performance while 
preserving the fine temporal features required for subsequent particle 
classification.

\subsection{Particle identification}\label{subsec:id}

\begin{table*}[t!]
\centering
\begin{tabular}{l c c c c}
\hline
\textbf{Model} & \textbf{Precision} ($\uparrow$) & \textbf{Recall} ($\uparrow$)
               & \textbf{Accuracy (prompt $\gamma$)} ($\uparrow$) & \textbf{MCE} ($\downarrow$) \\
\hline
\texttt{cnn\_fc\_100k} & 0.980 & 0.980 &
{0.911} & 0.220\\
\texttt{cnn\_fc\_100k\_calib} & 0.979 &
0.979 & \colorbox{green!20}{0.923} & \colorbox{green!20}{0.055}\\
\texttt{cnn\_vbll\_100k} & \colorbox{green!20}{0.981} &
\colorbox{green!20}{0.981} & 0.907 & 0.207\\
\hline
\texttt{cnn\_fc\_10k} & {0.978} & 0.978 & {0.915} &
{0.230}\\
\texttt{cnn\_vbll\_10k} & {0.980} & {0.980} & {0.908} & 0.315\\
\hline
\end{tabular}
\caption{Evaluation metrics on test set for five selected model variations.
Metrics reported are precision, recall, accuracy on prompt $\gamma$ subset, and
maximum calibration error (MCE). Precision and recall are respectively defined as $P =
\frac{\mathrm{TP}}{\mathrm{TP} + \mathrm{FP}}$ and
$R=\frac{\mathrm{TP}}{\mathrm{TP} + \mathrm{FN}}$, where $\mathrm{TP}$ are true
positives, $\mathrm{FP}$ are false positives, and $\mathrm{FN}$ are false
negatives. Best values are highlighted.}
\label{table:eval_metrics}
\end{table*}

\begin{figure*}[t!]
    \centering
    \includegraphics[width=0.8\linewidth]{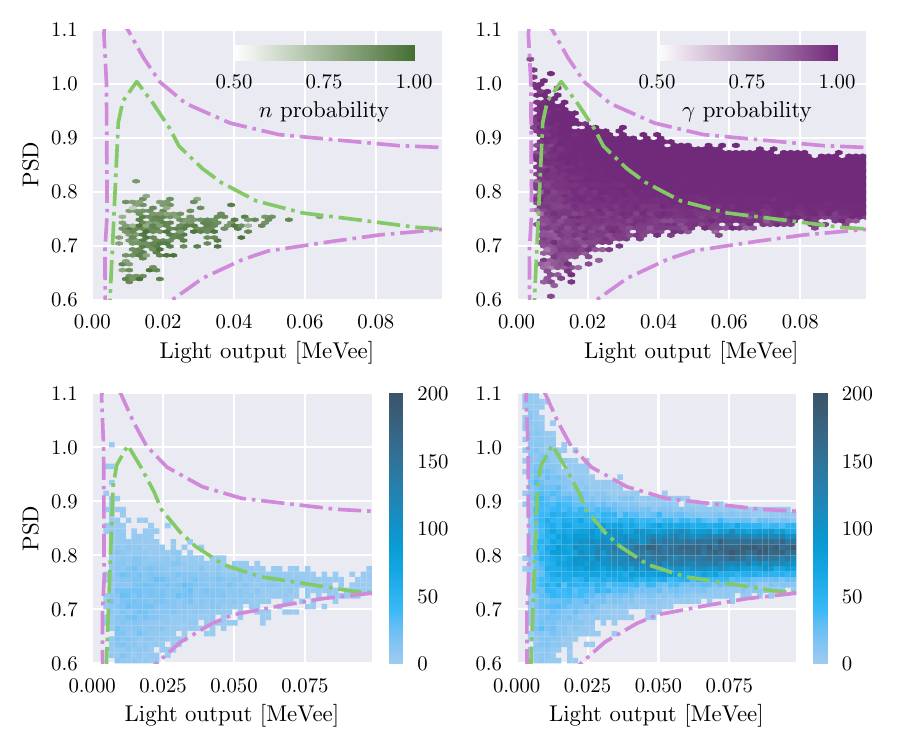}
    \caption{PSD as a function of the light output for low-energy prompt
    $\gamma$ rays ($0.8 < \beta < 1.2$). Left panels show predicted neutrons and
    right panel shows predicted $\gamma$ rays. Top panels: Bin color represents average probability
    computed for the bin (containing at least 8 events). Bottom panels: Bin
    color represents event distribution. 3$\sigma$-envelopes from~\cite{syrett2025vendeta} are
represented for the neutron and $\gamma$ clusters.}
    \label{fig:dea_prompt_perf}
\end{figure*}

The particle identification performance of the proposed framework is evaluated
on the test set, with quantitative results summarized in
Table~\ref{table:eval_metrics}. We benchmark five model configurations trained
with either 50{,}000 or 5{,}000 samples per class, and using either a standard
fully-connected classification head or a VBLL.
The model predictions exhibit very good agreement with the PSD method in the
low-charge region that was not used for training, with $P=0.981$ and $R=0.981$
for \texttt{cnn\_vbll\_100k}. This demonstrates strong extrapolation capabilities and indicates that the learned latent representation captures physically meaningful features of neutron and $\gamma$ waveforms beyond the charge range observed during training.
Both fully-connected (FC) and VBLL-based classifiers achieve comparable
precision and recall across all training set sizes. For models trained on
50,000 samples per class, an accuracy of 92.3\% is obtained on the prompt
$\gamma$ rays subset for \texttt{cnn\_fc\_100k\_calib}, providing an independent and 
high-purity validation of the classification performance.
Fig.~\ref{fig:dea_prompt_perf} shows the PSD factor as a function of the light 
output for the low-energy prompt $\gamma$ rays selected using ToF information. 
The bin color represents the average predicted probability for the bin. We get $7.7\%$
of false negative errors for charges under 100 keVee, which correspond to neutron predictions with
the \texttt{cnn\_fc\_100k\_calib} model variant. We stress that those neutron 
identifications are observed at the interface with the neutron cluster in the 
PSD space for low light output values, where the traditional PSD method alone 
fails to provide reliable classification. 
We observe that signals classified as 
neutrons by our model, using a threshold of 0.5, are associated to lower probability values 
than those classified as $\gamma$ rays. When setting a threshold on the neutron
and $\gamma$ probability to 0.999, we identify 76\% of the prompt $\gamma$ rays
as $\gamma$ and 0.5\% as neutrons, leaving 23.5\% of signals for which we do
not provide a label.
We show that the distribution of prompt $\gamma$-ray events labeled as 
neutrons by our model with a threshold of 0.5 falls within the $\gamma$ cut 
provided in~\cite{syrett2025vendeta} (bottom-left panel).
This, combined with the ToF argument, indicates 
that these signals correspond to $\gamma$ rays, therefore confirming that 
these events are indeed mislabeled by our model.
When reducing the training set size by an order of magnitude, the drop in 
precision and recall remains below 1\%.

\begin{figure}[t!]
    \centering
    \includegraphics[width=1.0\linewidth]{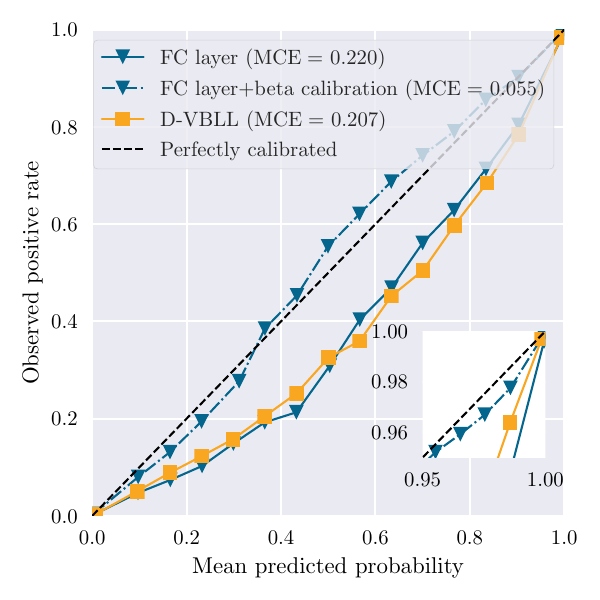}
    \caption{Calibration curves for the models trained on 50{,}000 samples per
    class using a final fully-connected layer (blue lines with triangle
  markers) and variational Bayesian last layer (orange line with square
markers). The black dashed line would correspond to a perfectly calibrated
classifier.}
    \label{fig:calib}
\end{figure}

We also evaluate the maximum calibration error (MCE) to quantify the agreement
between predicted probabilities and accuracy on identifying events
from an independent, augmented validation set, that is to assess the 
reliability of our predictions. 
Intuitively, a well-calibrated model should be correct about 80\% of the time
($\mathrm{acc} = 0.8$)
when it reports 80\% probability. MCE quantifies deviations from this ideal 
behavior by grouping predictions into $M$ confidence bins and comparing, within 
each bin, the average predicted probability to the model accuracy. 
It is defined as~\citep{pavlovic2025understanding}
% \begin{equation}
% \mathrm{ECE} = \sum_{m=1}^M \frac{|B_m|}{N}|\mathrm{acc}(B_m) - \mathrm{conf}(B_m)|,
% \end{equation}
\begin{equation}
  \mathrm{MCE} = \max_m|\mathrm{acc}(B_m) - \bar{p_n}(B_m)|,
\end{equation}
where $B_m$ denotes the set of predictions in bin $m$, $|B_m|$ is the number of
samples in that bin, and $N$ is the total number of samples. 
$\mathrm{acc}(B_m)$ and $\bar{p_n}(B_m)$ correspond to the 
accuracy and the mean predicted neutron probability within bin $m$, respectively. 
Lower MCE values indicate better-calibrated probabilistic predictions. 
Calibration curves on an independent, augmented validation set are also 
provided in Fig.~\ref{fig:calib}. The curves for
\texttt{cnn\_fc\_100k} (solid blue line with triangle markers) and
\texttt{cnn\_vbll\_100k} (solid orange line with square markers) show 
moderate deviations from the ideal diagonal with corresponding $\mathrm{MCE}$
values of $0.220$ and $0.207$, respectively.
We find that the calibration level of VBLL models is comparable to that of FC
models, with only minor variations across dataset sizes. This indicates that 
neither approach demonstrates a clear calibration advantage on this dataset.
Furthermore, VBLL is a custom layer that precludes exporting the trained 
network to a framework-agnostic format such as ONNX~\citep{onnx}, which is recommended for 
production deployments~\citep{araz2024houches}, making the FC-based model the 
more practical choice.
Nevertheless, the VBLL formulation provides a principled approach to 
uncertainty quantification which could be explored in future work.
%
% something on beta calibration for post-hoc calibration
We also experimented with post-hoc calibration
techniques~\citep{wang2023calibration}, among which beta
calibration~\citep{kull2017beyond} (dash-dotted line with triangle markers)
proved most effective. This method rescales the classifier's output scores
using a small set of parameters fit on a held-out validation dataset used for
calibration, correcting for systematic biases in the predicted probabilities 
without modifying the classifier itself.
As shown in Fig.~\ref{fig:calib}, applying beta calibration yields better-calibrated 
probabilities ($\mathrm{MCE} = 0.055$) without degrading classification performance (see
\texttt{cnn\_fc\_100k\_calib} row in Table~\ref{table:eval_metrics}).

\begin{figure}[t!]
    \centering
    \includegraphics[width=\linewidth]{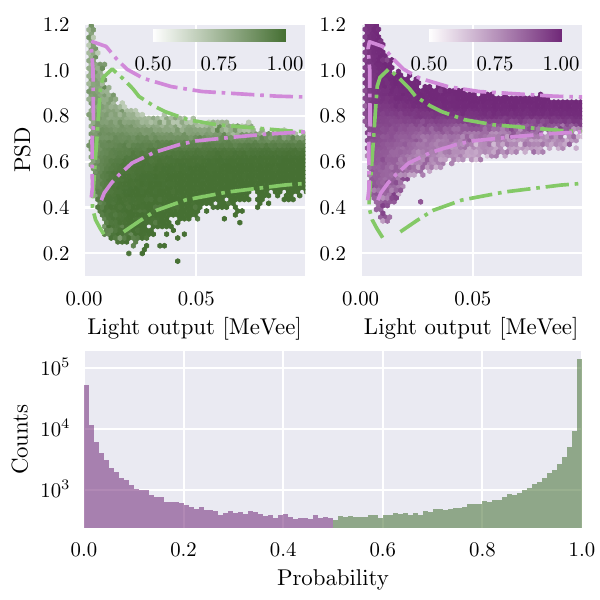}
    \caption{PSD as a function of light output in MeVee with probabilities on
    test set for predicted neutrons (left) and predicted $\gamma$ (right) using
  \texttt{cnn\_vbll\_100k}. Bin color represents the average probability
for at least 4 events. 3$\sigma$-envelopes from~\cite{syrett2025vendeta} are
represented for the neutron and $\gamma$ clusters. Bottom panel shows corresponding test set probability distribution.}
    \label{fig:predicted_w_unidentified}
\end{figure}

\begin{figure}[t!]
  \centering
  \includegraphics[width=1.0\linewidth]{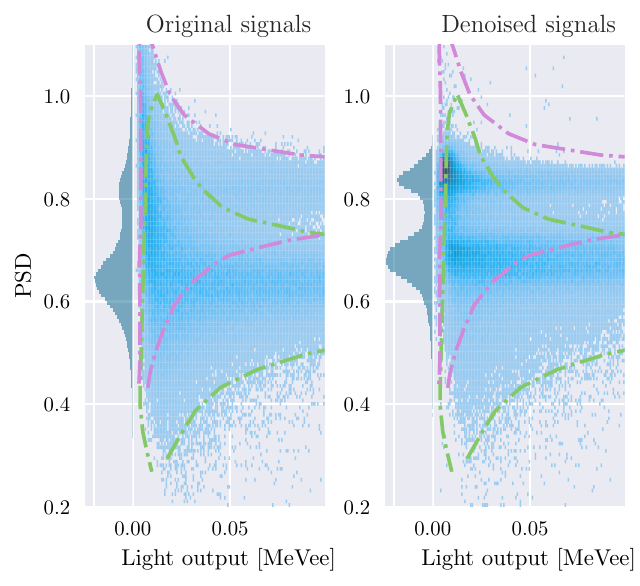}
  \caption{Density matrix of the PSD factor as a function of light output in
  the low charge regime (between 6 and 100 keVee) for original signals (left)
  and signals denoised with \texttt{cnn\_fc\_100k\_calib} (right). 3$\sigma$-envelopes from~\cite{syrett2025vendeta} are
represented for the neutron and $\gamma$ clusters.}
  \label{fig:denoised_psd}
\end{figure}

Fig.~\ref{fig:predicted_w_unidentified} shows the PSD factor as a function of
light output with predicted neutron (left panel) and $\gamma$ probabilities
(right panel) using the \texttt{cnn\_vbll\_100k} model variant below 100 keVee,
where graphical cuts cannot provide reliable labels. Note that a threshold 
value of 0.5 is used to distinguish between the two classes.
For both detected neutron and $\gamma$ events, we can see that the probability
slightly decreases with decreasing light output, and that the average predicted
probability reaches its lowest values at the interface between the neutron and
gamma distributions. Still, our models achieve a clear separation at low
charge, as highlighted by the bottom panel showing the predicted probability distributions
on the test set, and allow going beyond the graphical-cutting method in this 
regime.
This can be partly attributed to the fact that our encoder is trained to 
eliminate high frequency noise, therefore pushing the separability limit below
100 keVee as we can see by comparing the two panels of
Fig.~\ref{fig:denoised_psd}, which shows that the traditional PSD method is 
limited by the SNR level.

\begin{figure}
  \centering
  \includegraphics[width=1.0\linewidth]{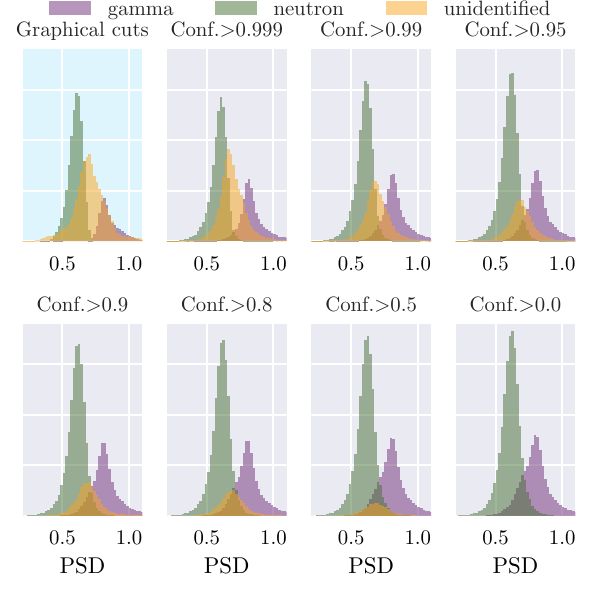}
  \caption{PSD factor distributions for signals with light output below 100
    keVee for multiple selected confidence $|p_n - p_\gamma|$ levels with
    \texttt{cnn\_fc\_100k\_calib}. 
    Green distributions correspond to predicted neutrons, purple 
    distributions correspond to $\gamma$, and unidentified signals are 
    represented by orange distributions. First panel distributions are obtained
  with graphical cuts~\citep{syrett2025vendeta}.}
  \label{fig:mean_margin_conf}
\end{figure}

Since our \texttt{cnn\_fc\_100k\_calib} model variant is well calibrated, we 
can reasonably define the margin confidence as $|p_n - p_\gamma|$ in order to 
quantify classification certainty. One could set a threshold on the confidence
value to fit their use case. In the case where a model prediction would be
associated with a confidence below threshold, the signal would be considered as
unidentified. Typically, for prompt fission neutron measurements, where
$\gamma$-ray exclusion is crucial, one could set the threshold value as high as 0.999.
Fig.~\ref{fig:mean_margin_conf} and Table~\ref{table:confidence} shows the
evolution of neutron, $\gamma$, and unidentified distributions with different confidence levels.

\begin{table*}
  \centering
  \begin{tabular}{l c c c c}
    \hline
    \textbf{Model} & \textbf{Confidence thr.} & \textbf{$n$ pop.~[\%]} &
    \textbf{$\gamma$ pop.~[\%]} & \textbf{Unidentified~[\%]} \\
    \hline
    Graphical cuts & -- & {41.6} & {12.9} & {45.5}\\
    \hline
    \texttt{cnn\_fc\_100k\_calib} & {0.999} & {41.9} & {21.2} & {36.9}\\
    \texttt{cnn\_fc\_100k\_calib} & {0.99} & {49.3} & {25.1} & {25.6}\\
    \texttt{cnn\_fc\_100k\_calib}& {0.95} & {53.8} & {27.9} & {18.3}\\
    \texttt{cnn\_fc\_100k\_calib} & {0.9} & {55.6} & {29.4} & {15.0}\\
    \texttt{cnn\_fc\_100k\_calib}& {0.8} & {57.3} & {31.1} & {11.6}\\
    \texttt{cnn\_fc\_100k\_calib}& {0.5} & {59.9} & {34.0} & {6.1}\\
    \texttt{cnn\_fc\_100k\_calib}& {0.0} & {62.8} & {37.2} & {0.0}\\
    \hline
  \end{tabular}
  \caption{Evolution of neutron, $\gamma$, and unidentified populations with
  confidence $|p_n - p_\gamma|$ level for test set signals with light output 
below 100 keVee. First row corresponds to the populations as obtained with 
graphical cuts~\citep{syrett2025vendeta}.}
  \label{table:confidence}
\end{table*}

\section{\label{sec:interpretability}Model explainability}

\begin{figure}
    \centering
    \includegraphics[width=0.99\linewidth]{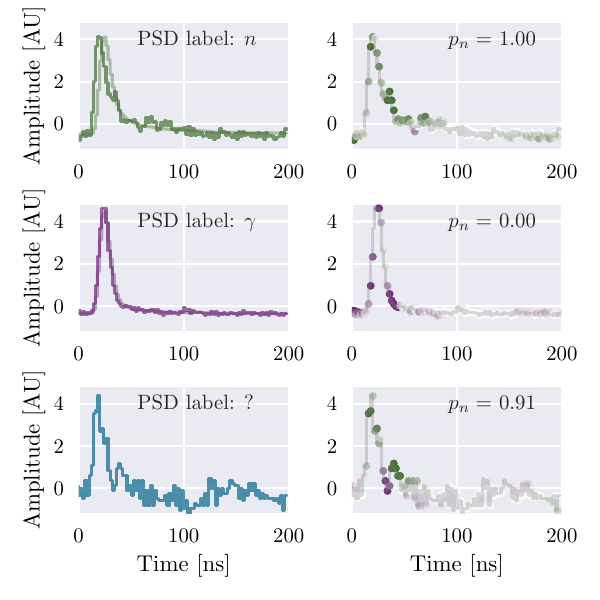}
    \caption{SHAP values for three signals. Left: Signal with its VENDETA PSD label. Average neutron and $\gamma$-ray waveforms are also shown for reference. Right: SHAP values and predicted neutron probability. Green (purple) markers indicate features that increase neutron ($\gamma$-ray) probability. The darker the color, the larger the SHAP value magnitude.
    }
    \label{fig:saliency_a}
\end{figure}

\begin{figure}
    \centering
    \includegraphics[width=0.99\linewidth]{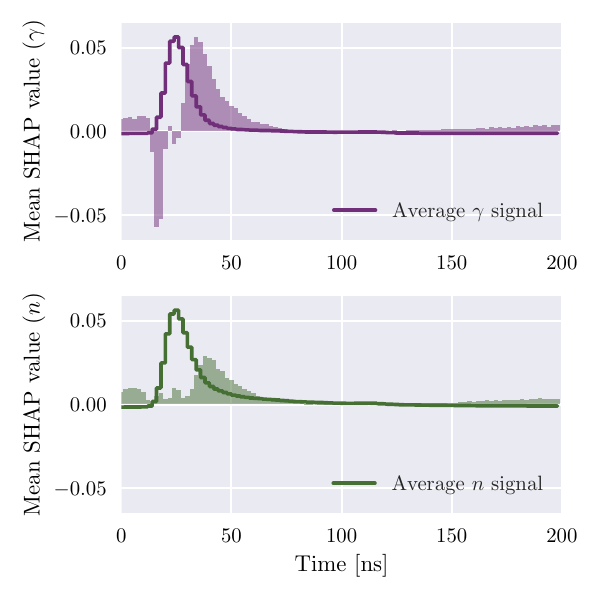}
    \caption{Mean SHAP value as a function of time for predicted $\gamma$-ray 
      (top) neutron (bottom) signals.}
    \label{fig:saliency_b}
\end{figure}

% what it model interpretability
Model explainability describes the ability to understand and explain how a trained model makes its decisions~\citep{longo2024explainable}, in our case why a signal would be associated to either a $\gamma$ or a neutron identification.
% some words on why this is important
According to~\cite{doshi2017towards}, the demand for interpretability stems from the inherent incompleteness of problem formalization. In such settings, it is necessary not only to know what the model predicts, but also to understand why it produces a given prediction, since correctness by itself only partially addresses the underlying problem.
In this work, the use of explainability techniques is motivated by both safety considerations and the need for scientific understanding.
% the safety perspective
From a safety perspective, we want to make sure that neutron-$\gamma$ discrimination decisions are driven by physically meaningful differences in pulse shape characteristics rather than spurious correlations or artifacts in the data.
% the scientific understanding perspective
From a scientific standpoint, explainability provides insight into which aspects of the waveform drive the model’s decisions, potentially guiding improvements in data acquisition and analysis methodologies.

% we propose to use SHAP values to interpret our model decisions. we prefer using SHAP instead of gradient-based methods
To interpret the decision-making process of our models, we employ SHAP (SHapley Additive exPlanations) values~\citep{lundberg2017unified}. SHAP provides a theoretically grounded framework for attributing a model’s prediction to individual input features based on Shapley values from cooperative game theory, ensuring properties such as consistency and local accuracy. We favor SHAP over gradient-based saliency methods, which have been shown to lack robustness in some scenarios~\citep{adebayo2018sanity,kindermans2019reliability}.
% how to interpret SHAP values
SHAP values quantify the contribution of each input feature to a specific prediction by measuring how the model output changes when that feature is included or excluded relative to a reference baseline. Positive SHAP values indicate that a feature increases the likelihood of a given class assignment, while negative values indicate a suppressing effect. In this work, SHAP analysis enables a physically interpretable assessment of which regions of the pulse waveform or derived signal features drive the classification of events as neutrons or $\gamma$ rays.

% introduce our figure
Fig.~\ref{fig:saliency_a} illustrates SHAP values computed using the enhanced DeepLIFT algorithm~\citep{shrikumar2017learning}, as implemented in the \texttt{shap} Python library~\citep{NIPS2017_7062}.
The left column shows three representative input waveforms with their
corresponding PSD labels: neutron (upper panel), $\gamma$-ray (center panel), 
and unidentified (lower panel). Faint lines represent the average neutron and 
$\gamma$-ray signals.
The right column display the approximate SHAP values with the predicted
neutron probability $p_n$ reported in each panel.
Note that since we are solving a binary classification problem we have $p_\gamma = 1 -
p_n$ and SHAP values are complementary: features that contribute positively to 
the probability of one class (here neutron) necessarily contribute negatively 
to the other (here $\gamma$ rays).

% what do we get from our results
The SHAP attributions indicate that the model primarily bases its
neutron-$\gamma$ discrimination on the beginning of the tail.
For events classified as $\gamma$ rays (middle row), positive SHAP values
for the $\gamma$-ray class are concentrated at the beginning of the tail, and
small peaks in the rest of the tail contribute negatively.
Conversely, for neutron-classified events (top row), we get high SHAP values 
when a high amplitude is observed at the beginning of the tail, which is 
consistent with the larger slow-component contribution expected from neutron events.
% ambiguous case
In ambiguous cases, where predicted class probabilities are closer to $0.5$,
the SHAP values are more distributed across the waveform, indicating competing
contributions from different signal regions. This behavior reflects the limited 
separability in the overlap region of the PSD space and aligns with physical 
limitations of neutron-$\gamma$ discrimination at low light output.
% right column
Finally, the mean SHAP values averaged across all predicted $\gamma$-rays and 
neutron signals are represented in Fig.~\ref{fig:saliency_b},
respectively, with the average signal overlaid.
% driving region at the beginning of the tail, between 25 and 75ns
It is seen that the most discriminative region lies in the tail of the pulse, 
between 25 and 75 ns, where mean SHAP values reach their largest absolute 
magnitudes, which is consistent with the fact that neutron and $\gamma$-ray 
pulses differ primarily in their tail decay time.
% low contribution beyond 75ns, indicating that we could truncate the signals
Beyond 75 ns, mean SHAP values are near zero for both classes, suggesting that 
the latter portion of the signal carries negligible information regarding
discrimination. This means that the signals could be safely truncated at this point 
without significant loss of performance.
The large negative SHAP contribution observed around the rising edge of the predicted $\gamma$-ray pulses is not fully understood. This part of the waveform suppresses the predicted $\gamma$-ray probability, thereby favoring the neutron class. A possible interpretation is that the model uses the early part of the pulse as an initial indication of the class and subsequently adjusts its prediction based on the later pulse evolution.

% summary
Overall, this analysis demonstrates that the model's decisions are driven by
physically meaningful pulse-shape features rather than isolated or spurious
signal fluctuations. It confirms that the learned representations are 
consistent with established PSD principles.

\section{\label{sec:conclusions}Conclusions}

% summary
In this work, we propose a lightweight deep learning framework for accurate and
explainable neutron–$\gamma$ discrimination in the low-charge regime, where
traditional pulse-shape discrimination methods fail to provide a clear
separation.
Data obtained in the detection of neutrons and $\gamma$ rays emitted following 
the spontaneous fission of $^{240}\text{Pu}$ with VENDETA were used as input to 
a network trained in a multitask setting with a shared encoder jointly 
optimized for signal denoising and particle classification.
This design encourages the learned representations to capture 
both the global pulse morphology and the subtle shape differences required for 
reliable discrimination, while limiting the size of the required training 
dataset. Because PSD labels become unreliable at low charge, training is 
restricted to events with light output above 200~keVee, and random data 
augmentations are applied to emulate lower signal-to-noise conditions. 
We demonstrate that the proposed model achieves a compression ratio of 12.5 
while providing superior denoising performance compared to conventional digital 
signal processing techniques, including total-variation denoising and low-pass 
filtering.
In terms of particle identification, a key contribution of our work is the 
transformation of the conventional two-dimensional graphical cut approach into 
a continuous confidence score. Rather than assigning a class label based on a 
fixed boundary, our model outputs a calibrated classification probability that 
allows the user to define the acceptable level of cross-contamination in a 
given region of the charge spectrum.
This flexibility is particularly valuable in the low-charge regime, below
100~keVee, where graphical cuts yield overlapping boundaries. 
In this region, our best performing model identifies 63.1\% of test-set signals as neutrons or $\gamma$ rays at a 99.9\% confidence level compared to 55.5\% recovered by graphical cuts. For a 99\% confidence level of our model, 74.4\% of the test-set signals are identified.
On a high-purity prompt $\gamma$-ray subset, our model achieves 92.3\% accuracy. 
These outcomes are grounded in 
physically meaningful pulse-shape features, as confirmed by SHAP analysis, 
which shows that model decisions are driven by the same temporal regions 
of the waveform that underpin established PSD principles rather than 
by isolated or spurious signal fluctuations.

While training deep learning models typically requires dedicated software
environments, inference should
integrate seamlessly into traditional analysis pipelines. To this end, an
\emph{nptool v4} plugin~\citep{matta2016nptool}, \emph{npsinapse}, was 
developed to perform CPU-based inference, in keeping with the event-by-event
analysis paradigm common to most nuclear physics experiments. 
The model is exported to the framework-agnostic ONNX format~\citep{onnx}, and
\emph{npsinapse} serves as a thin wrapper around the ONNX Runtime C++
API~\citep{onnxruntime}. Going forward, \emph{npsinapse} is intended to evolve into a 
general-purpose, templated interface to ONNX Runtime, enabling straightforward 
deployment of deep learning models within the \emph{nptool} ecosystem.
We also publish a dataset of preprocessed, labeled signals collected with the VENDETA array that we hope can serve as a reference benchmark for future studies on neutron-$\gamma$ discrimination~\citep{owen2026benchmark}. The dataset consists of four subsets: \emph{train}, \emph{val}, \emph{test}, and \emph{prompt\_gammas}.

% outlooks
While our results are encouraging, many challenges remain.
One is to explore the applicability of our approach for multiclass 
pulse-shape-analysis-based particle discrimination on GRIT data~\citep{grit}.
Another challenge is to optimize model inference for deployment on the next 
generation of the FASTER data acquisition system~\citep{faster} through 
techniques such as model quantization and TensorRT~\citep{tensorrt}, with an 
aim to provide efficient online data analysis. 
Indeed, beyond its role as an auxiliary training objective, the denoising 
branch serves practical purposes in its own right as the compressed latent 
representation significantly reduces data storage and transmission 
requirements.
Finally, we plan to investigate semi-supervised 
pre-training approaches, such as MixMatch~\citep{berthelot2019mixmatch}, to 
better leverage non-annotated data and reduce dependence on high-quality 
labels, particularly in low-charge regimes.

\section*{Acknowledgements}

Discussions with Franck Delaunay from LPC Caen are gratefully acknowledged.
TC, AM, and DE acknowledge the financial support of the SINAPSE grant from
Normandie Recherche. This work  was supported by the US Department of Energy through the
Los Alamos National Laboratory, United States. Los Alamos National
Laboratory is operated by Triad National Security, LLC, for the National
Nuclear Security Administration of US Department of Energy (Contract
No.~89233218CNA000001).

\section*{Data Availability}

Our training framework, \emph{sinapse-training}, is openly available at \url{https://gitlab.in2p3.fr/sinapse/sinapse-training} and ONNX model weights for \texttt{cnn\_fc\_100k\_calib} are part of the SINAPSE \emph{nptool} plugin available at \url{https://gitlab.in2p3.fr/sinapse/npsinapse}.

We publish a dataset of preprocessed, labeled signals collected with the VENDETA array that can serve as a reference benchmark for neutron-$\gamma$ discrimination studies~\citep{owen2026benchmark}. It comprises four subsets: \emph{train}, \emph{val}, \emph{test}, and \emph{prompt\_gammas}.

\bibliography{apssamp}% Produces the bibliography via BibTeX.

\end{document}